\documentclass[showpacs,preprintnumbers,amsmath,amssymb, superscriptaddress]{revtex4}

\usepackage{graphicx}
\usepackage{dcolumn}
\usepackage{bm}
\usepackage[usenames]{color}    

\newcommand{\be}{\begin{equation}}
\newcommand{\ee}{\end{equation}}
\newcommand{\bea}{\vspace{0.25cm}\begin{eqnarray}}
\newcommand{\eea}{\end{eqnarray}}

\def\PRL{{Phys. Rev. Lett.} }
\def\PRA{{Phys. Rev.} A }

\begin{document}
\title{Systematic analysis of SNR in bipartite Ghost Imaging with classical and quantum light }
\author{G.Brida $^1$,M.~V.~Chekhova$^{1,2,4}$, G.A. Fornaro $^{1}$, M.~Genovese$^1$, L.Lopaeva$^{1,3}$, I. Ruo Berchera}

\affiliation{
 Istituto Nazionale di Ricerca Metrologica, Strada
delle Cacce 91, 10135 Torino, Italy\\ $^2$Max-Planck Institute for
the Science of Light, Staudtstrasse 7, 91058 Erlangen, Germany \\
$^3$ Dipartimento di Fisica, Politecnico di Torino, I-10129, Torino,
Italy \\ $^4$Department of Physics, M.V.Lomonosov Moscow State
University, Leninskie Gory, 119992 Moscow, Russia}
\begin{abstract}
We present a complete and exhaustive theory of
 signal--to-noise--ratio  in bipartite  ghost imaging  with classical
(thermal) and quantum (twin beams) light. The theory is compared
with experiment for both twin beams and thermal light in  a certain regime of interest.
\end{abstract}
 \maketitle
\section{Introduction}\label{}

Spatial optical correlations, both quantum and classical, represent
a fundamental resource for developing technologies as quantum
imaging \cite{l1,k}, super-resolution \cite{sr}, etc., which could
open unprecedented opportunities in the field of metrology,
positioning and imaging. Various protocols have been proposed
\cite{th2,th3,th5,th6,th7,th8,th9,th10,th11,th12} and experimentally
realized
\cite{exp1,exp2,exp4,exp5,exp6,exp7,exp8,exp9,exp10,BridaPRL2009,BridaNPHOT2010,BDGRR-OE-2010}.

Among them a great interest has been attracted by the so called
ghost imaging \cite{gh1} (GI), based on the correlation in
 spatial  intensity
fluctuations (speckles \cite{sp}). In this technique a light beam
crosses (or it is reflected)  by an object to be imaged.
 However, the beam that crossed the object is
detected by a detector without any spatial resolution (bucket
detector).  The image of the object is retrieved when the bucket
detector signal is correlated with the signal of a  spatially
resolving detector measuring a light beam whose noise is spatially
correlated to the previous beam (reference beam). The first
demonstration of this technique was achieved with nonclassical
states of light \cite{gh2}, known as twin beams, produced by
parametric down conversion (PDC). Then it was shown, both
theoretically and experimentally, that this result can be achieved
also with beam-split thermal light
\cite{g,gh31,gh32,gh33,gh34},  although  with a smaller visibility.

In practice this technique can be useful in the presence of phase
distortions (for example when the beam crosses a diffusive medium,
like fog), where intensity correlations with a second beam allow
one to retrieve the spatial information\cite{r}.

Due to the conceptual and practical interest attracted to GI, a lot
of works have been devoted to clearly theoretically describe
\cite{ESpra2009,exp6,per,m} and to improve this protocol
\cite{imp,g3,Chen,Teo,Dix}. The main parameters of GI discussed in
these works are typical for any kind of imaging: the signal-to-noise
ratio (SNR), which shows how well the image of the object is
distinguishable from the background, and the resolution. The latter
is related to the number of speckles (spatial modes) contained in
the image and is basically the number of elementary details of the
object that can be reconstructed in the ghost image. Note that
various terms are used in the literature:
SNR~\cite{Agafonov,ESpra2009,Basano} is sometimes called
contrast-to-noise ratio (CNR)~\cite{Chan2010}; number of spatial
modes~\cite{m,Agafonov} is sometimes called the number of details of
the image~\cite{Chen}.

Despite the huge amount of literature devoted to GI, several points
are still not clear concerning this technique. For instance, in all
existing experimental works, GI is performed by measuring either
normalized intensity correlation functions (CFs)
\cite{Agafonov,Chen} or intensity CFs with the background subtracted
(also called CFs of intensity fluctuations)~\cite{gh31,gh32,gh33}.
At the same time, up to recently, most theoretical papers considered
only the simplest CFs, without normalization or background
subtraction.  In a recent theoretical paper, Chan et
al.~\cite{Chan2010} for the first time considered both the previous
procedures and showed that they lead to almost the same SNR value,
while GI through the measurement of usual CFs has much smaller SNR.
However, they only considered GI with bright thermal light and
presented only the results of numerical simulations. In
\cite{ESpra2009} Erkmen et al. developed a theory of the SNR of GI
in the analog detection scheme using the not-normalized intensity
correlation, providing results when ac-coupling of the photocurrents
is performed by introducing  signal  frequency filtering.

In this paper,  for  providing the reader with a
general discussion of second-order~\footnote{we do not consider
here schemes based on higher-order correlation functions
\cite{th3,g3}.} GI in view of practical applications, we
generalize the theory of Ref~\cite{Chan2010} by including into
consideration, beyond thermal light ghost imaging (ThGI), the case
of twin beams (TwGI) and assuming an arbitrary brightness of the
light source. Moreover, we consider another protocol for GI that exploits the variance of the
difference signal from the object and reference detectors~\cite{m}.  It
turns out that this method of GI provides the same SNR as the
background subtraction and normalization of the CF for middle and high brightness of the source.

We analyze the influence of important aspects like the brightness of the source,
the losses, the number of spatiotemporal modes collected by  the detector, and the  resolution on the  performance  of GI in terms of SNR.

Finally, in this work we
compare the theory with the experiment, for two particularly interesting
cases: mesoscopic twin beams (photon numbers per spatiotemporal mode about $0.2$) and bright pseudo-thermal light, both detected by  CCD  cameras. The former is the best GI
option if high intensities should be avoided (for instance, for the
sake of not damaging the imaged object), and the latter is the
simplest way to achieve high SNR if using bright light is not
a problem.

\section{Theory}\label{Theory}

In general, in order to obtain a ghost image, a spatially incoherent
beam interacts with the object and then is collected by a bucket
detector without any spatial resolution, while the correlated beam,
which does not interact with the object,  is registered by a
spatially resolving detector, namely an array of pixels. This
procedure is repeated $\mathcal{K}$ times.  The image of the object
is described by  a parameter $S(x_{j})$, where $x_{j}$  represents
the position of the pixel $j$ in the reference region. Usually, $S$
has the  form
$$
S(x_{j})=f(E[\mathbb{N}_{1}],E[N_{2}(x_{j})],E[\mathbb{N}_{1}N_{2}(x)],E[\mathbb{N}_{1}^{2}],E[N_{2}^{2}(x_{j})],...),
$$
i.e., a function $f$ that involves the correlation function
$E[\mathbb{N}_{1}^{p}N_{2}^{q}(x_{j})],\,(p,q\geq0)$, of the observable corresponding to the total number of photons collected at the bucket detector and at the j-th pixel of the reference arm, respectively. Here, $E[X]= \frac{1}{\mathcal{K}}\sum^{\mathcal{K}}_{k=1} X^{(k)}$ represents the average over the set of $\mathcal{K}$  realizations.  For the sake of simplicity we consider an imaged object defined by two  levels  of transmission, $T=1$ and $T=0$.

The signal-to-noise ratio of a ghost imaging protocol can be
defined as the ratio  of the mean "contrast" to its standard deviation (mean fluctuation):

\begin{equation}
\label{SNRdef}
SNR_{S}\equiv\frac{|\left\langle S_{in}-S_{out}\right\rangle|}
{\sqrt{\left\langle\delta^{2}(S_{in}-S_{out})\right\rangle}},
\end{equation}
where $S_{in}$ and $S_{out}$ are the intensity values of the
reconstructed ghost image, when $x_{j}$ is either inside ($T=1$) or outside ($T=0$) of the object profile,  respectively, and \textbf{$ \delta S\equiv S-\langle S\rangle$}  is the fluctuation. The mean value, denoted by $\langle .\rangle$, represents the theoretical expectation value, which  can be estimated in practice  by performing space averages over the regions "in" and "out" of the ghost image.

The ghost image can be reconstructed by exploiting different GI protocols,
namely different parameters $S$. Although all the protocols exploit the correlations between the object beam and the reference one, not all off them give the same results in terms of SNR.  In particular,  in the following we analyze four different protocols based on using:

i) the Glauber intensity correlation function,
when $S(x)=G^{(2)}(x)\equiv E[\mathbb{N}_{1}N_{2}(x)]$.

Here, the quantum expectation value is obviously
$\langle S(x)\rangle=\langle\mathbb{N}_{1}N_{2}(x)\rangle$, and the standard deviation can be evaluated directly as $\langle\delta^{2}S(x)\rangle=\langle\delta^{2}[\mathbb{N}_{1}N_{2}(x)]\rangle/\mathcal{K}$. The  photon numbers are considered as quantum operators  and the quantum mean values are evaluated in the appendix, where we describe in detail our theoretical model.

ii) the normalized intensity CF, when
$S(x)=g^{(2)}(x)\equiv G^{(2)}/(E[\mathbb{N}_{1}]E[\mathbb{N}_{2}(x)])$.

We  note  that, unlike all the other parameters considered
at point i, iii and iv,  the standard deviation of $g^{(2)}$ measurement was evaluated in our work by means of the uncertainty propagation of  the quantities $G^{(2)}$, $E[\mathbb{N}_{1}]$ and $E[\mathbb{N}_{2}(x)]$.

iii) the covariance, or CF of intensity fluctuations,
$S(x)\equiv Cov(x)=E[(\mathbb{N}_{1}-E[\mathbb{N}_{1}])(N_{2}(x)-E[N_{2}(x)])]=E[\mathbb{N}_{1}N_{2}(x)]-E[\mathbb{N}_{1}]
E[N_{2}(x)]$.

In this case we have
$\langle Cov(x)\rangle=\frac{\mathcal{K}-1}{\mathcal{K}}(\langle\mathbb{N}_{1}N_{2}(x)\rangle-\langle\mathbb{N}_{1}\rangle\langle N_{2}(x)\rangle)$.
The coefficient depending on the number of  realizations,
 namely the number of acquired images, is the usual one allowing the unbiased estimation of the theoretical correlation. If we assume that $E[\mathbb{N}_{1}]$ and $E[\mathbb{N}_{2}]$ can be well represented by their expectation values $\langle\mathbb{N}_{1}\rangle$ and $\langle N_{2}\rangle$ in the calculation of the  fluctuations  (it happens if $\mathcal{K}\gg1$), we  have
$$
\langle \delta^{2} Cov(x)\rangle\simeq \langle\delta^{2}[\delta\mathbb{N}_{1}
\delta N_{2}(x)]\rangle/\mathcal{K}=(\langle[\delta\mathbb{N}_{1}\delta N_{2}(x)]^2\rangle-\langle\delta\mathbb{N}_{1}\delta N_{2}(x)\rangle^2)/\mathcal{K}.
$$
iv) the variance of the intensity difference,
$$
S(x)\equiv E[(\mathbb{N}_{1}-N_{2}(x)-E[\mathbb{N}_{1}-N_{2}(x)])^2]\equiv
(E[\mathbb{N}_{1}^2]-E[\mathbb{N}_{1}]^2)+(E[N_{2}(x)^2]-E[N_{2}(x)]^2)-2(E[\mathbb{N}_{1}N_{2}(x)]-E[\mathbb{N}_{1}]
E[N_{2}(x)]).
$$

Whatever the parameter $S$, it involves in some form the CFs,
which are second-order intensity moments. Therefore, the
evaluation of its  fluctuations  requires expressions for the first- to fourth-order moments of the intensity. These are calculated in the Appendix, where the details of our model are extensively
described.

We consider several parameters that influence the SNR.

\begin{itemize}
 \item The
brightness of the source, which is related to the average number of photons $\mu$ per single spatio--temporal mode.

  \item The total number $M$ of
spatio-temporal modes collected by each element (pixel) of the
spatially resolving detector in the reference arm. It is approximately
given by the product of the number of spatial modes,
$\mathcal{M}_{sp}=\max[\mathcal{A}_{pix}/\mathcal{A}_{coh},1]$, and
the number of temporal modes,
$\mathcal{M}_{t}=\max[\mathcal{T}_{det}/\mathcal{T}_{coh},1]$.
Here, $\mathcal{A}_{pix}$, $\mathcal{T}_{det}$ are the pixel size
and the detection (integration) time, respectively, while
$\mathcal{A}_{coh}$, $\mathcal{T}_{coh}$ represent the
characteristic coherence area (roughly, the speckle size) and the
coherence time of the source. From the viewpoint of the light
statistics, the total number of modes is $M\geq1$; it is considered
equal to 1 when $\mathcal{A}_{pix}\leq\mathcal{A}_{coh}$ and
$\mathcal{T}_{det}\leq\mathcal{T}_{coh}$. While through the
paper we always consider a usual situation in which the integration
time is much larger than the coherence time, the case of
$\mathcal{A}_{pix}\leq\mathcal{A}_{coh}$ and
$\mathcal{A}_{pix}>\mathcal{A}_{coh}$ are both analyzed in the text. We stress that $M$ should not be confused with the number of
realizations, for instance the number of acquired frames
$\mathcal{K}$:  $M$ is the number of modes collected by a pixel in a
single acquired frame,  so that $\mathcal{I}\equiv\langle
N_{2}\rangle=\eta_{2} M \mu$ represents the total number of photons
detected in the pixel,  further called the ``illumination level''.

  \item We take into account the overall transmission-collection-detection
efficiency of the two channels, $0\leq\eta_{j}\leq1$ with $j=1,2$,
i.e. the probability to detect an emitted photon. According
to our model, developed in the appendix, the situation in which the
pixel of the reference detector is smaller than the coherence area
($\mathcal{A}_{pix}\leq\mathcal{A}_{coh}$) corresponds to a reduction
of the collection efficiency. Therefore, $\eta_{2}$ includes also a
factor that takes into account for the geometrical collection
probability of a photon in a certain spatial mode,
$\eta_{2}=\eta_{2,0}*\eta_{2,coll}$, where
$\eta_{coll}\approx\min[\mathcal{A}_{pix}/\mathcal{A}_{coh},1]$.
  \item Finally, the number of spatial resolution cells of the
reconstructed image is represented by
$R=\mathcal{A}_{in}/\max[\mathcal{A}_{pix},\mathcal{A}_{coh}]$, with
$\mathcal{A}_{in}$ the area of the ghost image where $T=1$. When the
pixel size is equal or smaller than the size of a single spatial
mode, $R$ is determined by the number of spatial modes, otherwise it
is given by the number of pixels in the area $\mathcal{A}_{in}$.
The ideal condition for maximizing in principle the resolution and
the collection efficiency is
$\mathcal{A}_{pix}\simeq\mathcal{A}_{coh}$, that is, the pixel size
should approximately coincide with the speckle size \cite{k}.
\end{itemize}

As an example of a specific relevant case, in Tab. \ref{table} we
report the expressions for the SNR calculated for the ideal case of
unity transmission/quantum efficiency and collection
efficiency in both optical paths, $\eta_{1}=\eta_{2}=1$.  The SNR
values are normalized to $\sqrt{\mathcal{K}}$.

\begin{table}[h]
\centering
\begin{tabular}{||c|c|c||} \hline\hline
& \textbf{TwGI} & \textbf{ThGI} \\ \hline
$SNR_{G2}/\sqrt{\mathcal{K}}$ & $\frac{\sqrt{M \mu(1+\mu)}}{\sqrt{1+\mu (6+M+4 M R)+\mu2 \left(6+M+6 M R+2 M2 R2\right)}}$ &
$\frac{\sqrt{M} \mu}{\sqrt{1+2 M R+2 \mu \left(2+3 M R+M2 R2\right)+\mu2 \left(6+M+6 M R+2 M2 R2\right)}} $ \\ \hline $SNR_{g2}/\sqrt{\mathcal{K}}$ &
$ \frac{\sqrt{M R\mu(1+\mu)}}{\sqrt{1+\mu R (2+M+2 M R)+\mu2 \left(-1+(3+M) R+2 M R2\right)}}$ &
$\frac{\sqrt{M R}\mu}{\sqrt{-\mu (1+\mu)+\left(1+3 \mu+(3+M) \mu2\right) R+2 M (1+\mu)2 R2}}$ \\ \hline $SNR_{Cov}/\sqrt{\mathcal{K}}$ & $\frac{\sqrt{M \mu(1+\mu)}}{\sqrt{1+\mu (6+M+2 M R)+\mu2 (6+M+2 M R)}} $ & $\frac{\sqrt{M} \mu}{\sqrt{1+2 M R+4 \mu (1+M R)+\mu2 (6+M+2 M R)}} $ \\ \hline
$SNR_{Var}/\sqrt{\mathcal{K}}$ & $\frac{\sqrt{2 M \mu(1+\mu)}}{\sqrt{1+\mu (6+4 M R)+\mu2 (6+4 M R)}} $ &
$\frac{\sqrt{2M} \mu^{3/2}}{\sqrt{1+\mu (7+M (2+4 R))+8 \mu2 (1+M R)+\mu3 (6+4 M R)}} $ \\ \hline\hline
\end{tabular}
\caption{ Expressions for the signal-to-noise ratio (SNR)  of the
reconstructed ghost image with thermal light (ThGI) and with twin beams (TwGI) for  different  protocols described in the text, in the lossless case ($\eta=1$).}
\label{table}
\end{table}

\begin{figure}[tbp]
\center
\includegraphics[width=10cm, height=7cm ]{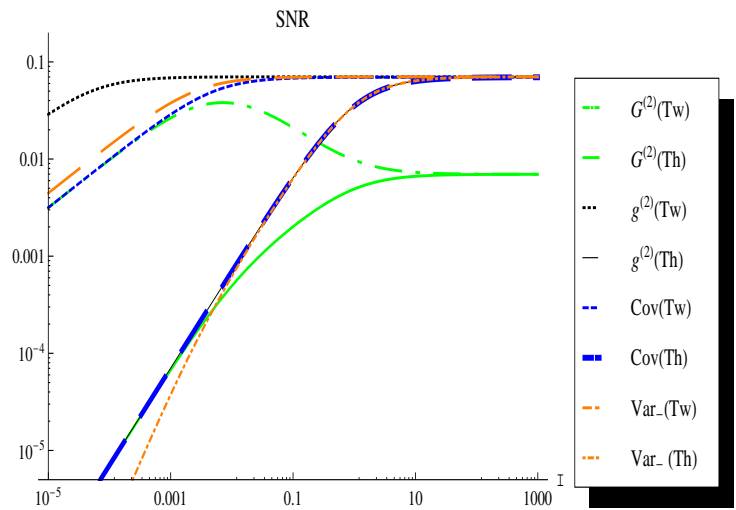}
\caption{ Signal-to-noise ratio  of the ghost image  as a function
of the illumination level $\mathcal{I}$ for different protocols
investigated in the paper. The number of detected spatio--temporal
modes in a  single run and the resolution parameter are fixed to be,
respectively, $M=1$ and $R=100$.} \label{SNRvsIsmall}
\end{figure}

First we observe that both for TwGI and ThGI the performances of the
scheme based on $G^{2}$ are drastically worse than for the other
methods because the SNR drops as $SNR_{G^2}\rightarrow1/R \,
(R\gg1)$ and as $1/\sqrt{M}\,(M\gg1)$ with respect to the resolution
parameter $R$ and with respect to the number of modes detected by
each pixel $M$.  At the same time, all the other protocols
(exploiting $g^{2}$, $Cov$ and $Var$) scale as $1/\sqrt{R}\,(R\gg1)$
and asymptotically $\sim \mathrm{const}$ for $M\gg1$. We report  the
theoretical dependence of SNRs on  $R$ in Fig. \ref{SNRvsRpdc} and
Fig.\ref{AllTogether} in Sec. \ref{Results}. This fact makes the
not-normalized and not-subtracted correlation function $G^{2}$,
albeit largely considered in the literature (see, for instance,
Refs.~\cite{ESpra2009,Basano,Chan2010}), inappropriate for ghost
imaging when compared with the other protocols. Thus, we do not
investigate it further through the paper.

Fig. \ref{SNRvsIsmall} presents the SNR for all the protocols  as a
function of the illumination level when $M=1$ and $\eta=1$. In the
typical situation of large $R$ ($R>10$) it turns out that
$SNR_{g^{2}}$, $SNR_{Cov}$ and  $SNR_{Var}$  behave the same for
large values of detected photons $\mathcal{I}\gg1$, approaching
approximatively the  upper value of $(2R)^{-1/2}$, although they
reach this bound at different values of the brightness. In
particular, for all ThGI protocols the bound is reached when
$\mathcal{I}\gg1$. For TwGI  based on the variance and covariance,
the condition depends on the resolution, namely $\mathcal{I}\gg 1/(2
R)$, while our calculation shows that $g^{(2)}$ reaches the flat
region as soon as $\mathcal{I}\gg1/( 2R^{2})$. Concerning the
performance of a very low-brightness  source, Fig. \ref{SNRvsIsmall}
shows that the  advantage  of using twin beams is in general very
pronounced but, even with the same source, some protocol seems to be
more convenient than others. In particular, for ThGI all the
protocols scale as $\propto\mathcal{I}$ with the exception of the
variance method, for which SNR  approaches zero faster, i.e.
$\propto\mathcal{I}^{3/2}$. For TwGI all the methods lead to the
same asymptotic behavior $\propto\mathcal{I}^{1/2}$. Therefore, we
can conclude that for the ghost imaging of a complex object the
three protocols exploiting $SNR_{g^{2}}$, $SNR_{Cov}$ and
$SNR_{Var}$ have the same  performance at medium and high
intensities (for the same source, quantum or classical), while  for
very low brightness, preference should be given to the normalized
correlation function $g^{2}$.

\begin{figure}[tbp]
\center
\includegraphics[width=10cm, height=7cm ]{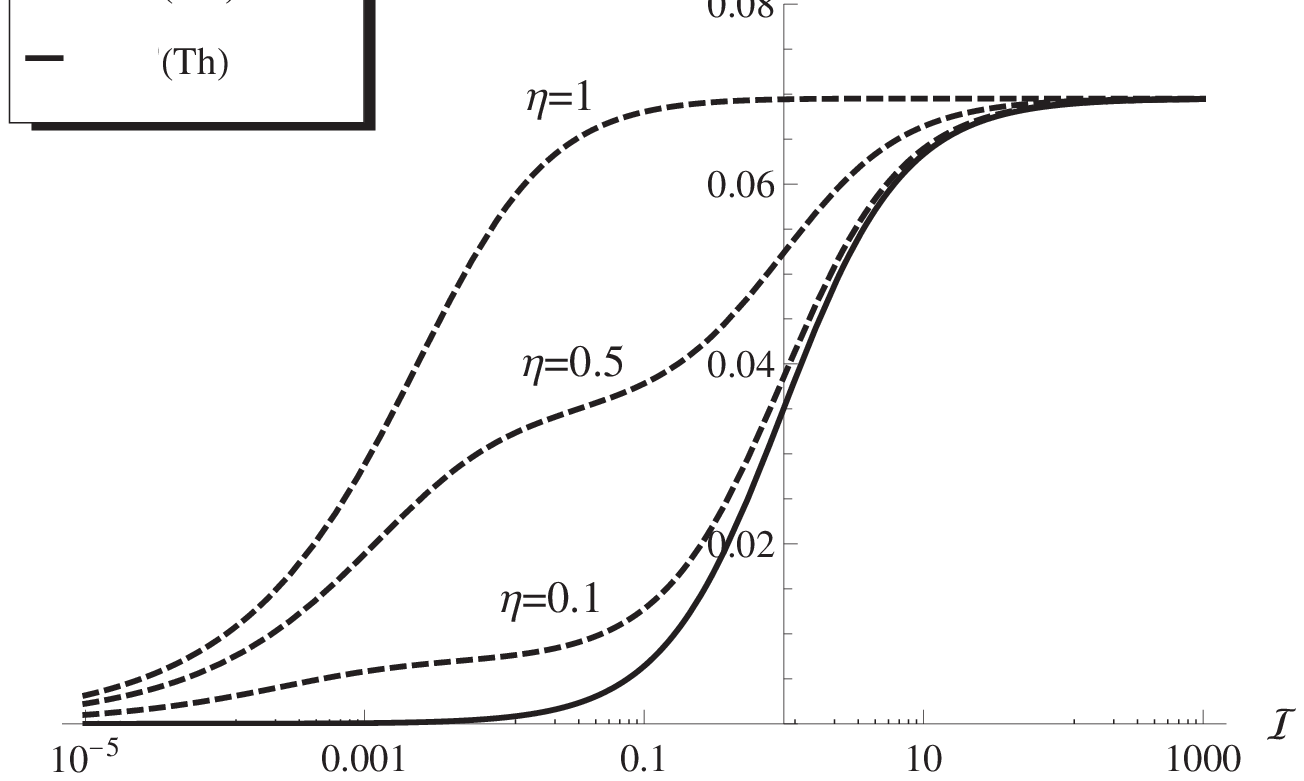}
\includegraphics[width=10cm, height=7cm ]{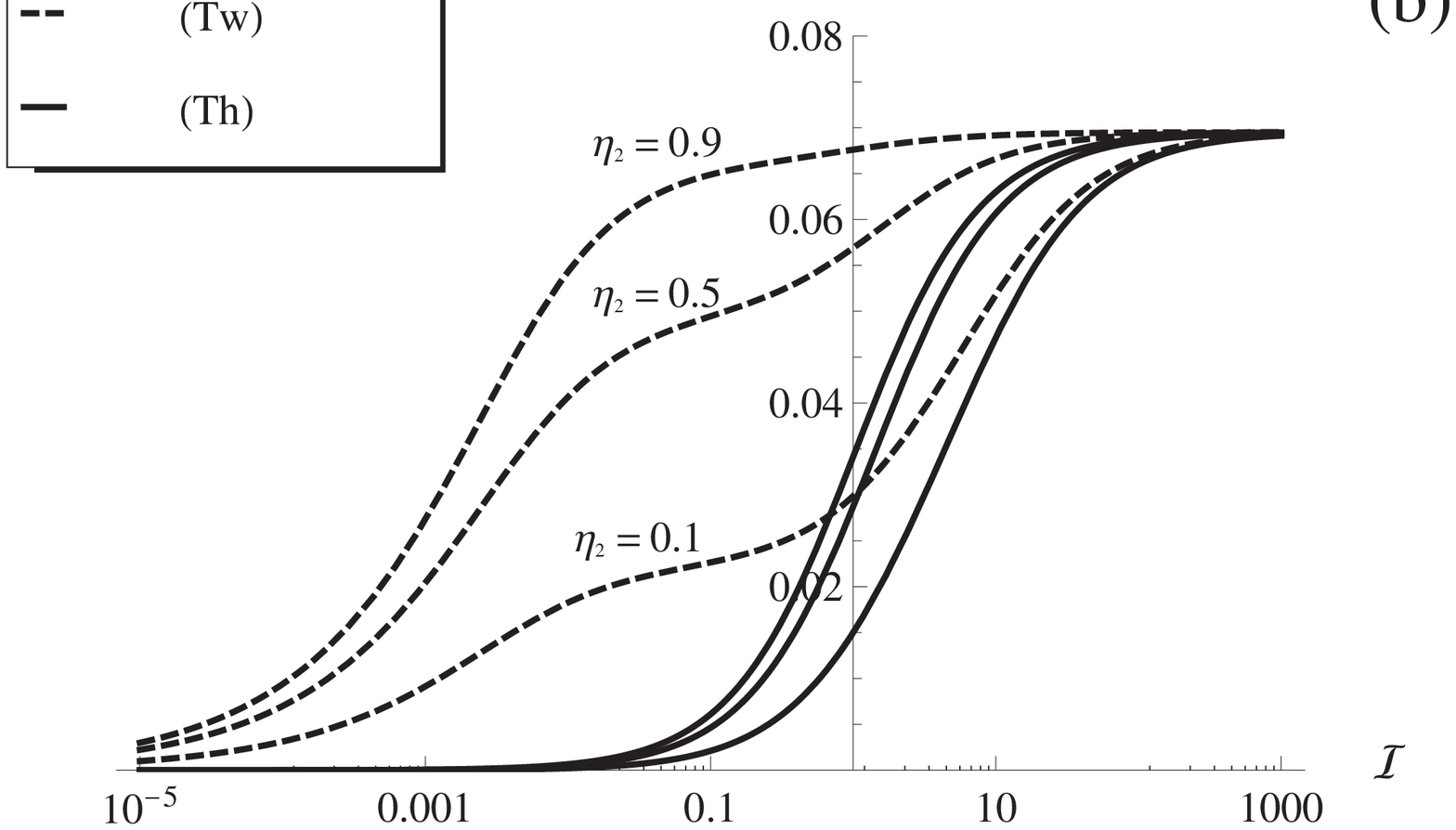}
\caption{ Signal-to-noise ratio  (SNR) of a ghost image  as a
function of the illumination level $\mathcal{I}$  for twin-beam and
thermal-light GI. The number of detected spatio--temporal modes in a
single run and the resolution parameter are fixed to be,
respectively, $M=1$ and $R=100$. (a): the three dashed curves refer
to different values of the  balanced detection probability
$\eta_{1}=\eta_{2}=\eta$. (b): the unbalanced case where
$\eta_{1}=1$ and $\eta_{2}=0.9,0.5,0.1$ from the higher to the lower
curve, respectively. } \label{SNRvsI-eta}
\end{figure}

Now we focus on the difference between the signal-to-noise ratio of
TwGI, $SNR_{Tw}$,  that is obtained in the limit of large R, and the
corresponding $SNR_{Th}$ obtained with thermal light in the same
limit,  by using the covariance  as the parameter for the GI
reconstruction. In order to perform a fair and probably more useful
comparison in a practical situation, we already introduced the
"illumination" $\mathcal{I}=\eta M \mu$, i.e. the total number of
photons detected in the pixel in the single image. In Fig.
\ref{SNRvsI-eta}(a) we report the quantum and classical SNR, for
different values of the transmission/detection efficiency, balanced
for the two channels, while fixing $M=1$ (a single temporal mode and
a single spatial mode are detected). Not surprisingly, the curve for
the thermal light is not influenced by the losses, while the TwGI
curve, in general showing a better performance than the classical
one, for high losses approaches the
 ThGI.  The advantage of twin-beam light is evident  for  a low-brightness source,  i.e. when the number of  photons  per
space-temporal mode $\mu\ll1$. On the contrary, when $\mu\gg1$,
thermal and  twin-beam states of light  produce the same results
(Fig.~\ref{SNRvsI-eta}). The physical reason can be understood
with the help of Eqs. (\ref{n-corr}), (\ref{n-crosscorr}) and
(\ref{n-crosscorrth}) for the two-mode  statistics. Both classical
and quantum light have the same single-mode thermal fluctuations
$\langle \delta^{2} n_{j}\rangle=\eta\mu+\eta^{2}\mu^{2}$.
However, the photon-number correlations for thermal light,
$\langle \delta n_{1}\delta n_{2}\rangle_{TH}=\langle
n_{1}\rangle\langle n_{2}\rangle=\eta^{2} \mu^{2}$, are relevant
when the number of photons is large,  $\langle
n_{j}\rangle\equiv\mu>1$, and do not  include the shot-noise
component $\propto\mu$, which remains uncorrelated. On the other
hand, the correlations in twin beams are $\langle \delta
n_{1}\delta n_{2}\rangle_{TB}=\eta^{2} \mu (1+\mu)$, which shows
that even the shot-noise component of the fluctuations  is
correlated. Moreover,classical two-mode correlations, generated by
a beam splitter obeying the Bernoulli statistics, are not reduced
by the losses, which are again described by a beam splitter model
(see the Appendix and the discussion after Eq.
(\ref{n-crosscorrth})).

It is also important to present (Fig.\ref{SNRvsI-eta}(b)) the
unbalanced case, $\eta_{1}>\eta_{2}$, which in our model includes
the typical situation in many experiments where the pixel is smaller
than the coherence area. In this case, $\eta_{2,
coll}\sim\mathcal{A}_{pix}/\mathcal{A}_{coh}\leq1$, and the
advantage of the TwGI is reduced. We observe that thermal ghost
imaging is weakly influenced by the unbalancing. This indicates
that, in order to make the best of the TwGI, in the photon-counting
regime one should properly set the pixel dimension.

\begin{figure}[tbp]
\center
\includegraphics[width=10cm, height=7cm ]{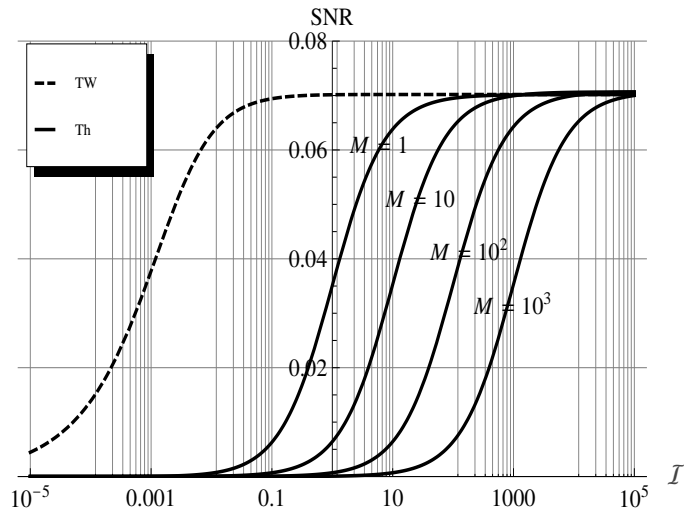}
\caption{Signal-to-noise ratio (SNR) of the ghost image as a
function of the illumination level $\mathcal{I}$. Here we
consider an  ideal lossless situation $\eta_{1}=\eta_{2}=1$ and the
resolution is $R=100$. }\label{SNRvsI-M}
\end{figure}

In Fig.\ref{SNRvsI-M}, the dependence of SNR on the number of
modes $M$ collected by a single pixel is shown. We see that the
 TwGI  is insensitive to $M$ while the SNR for  ThGI  is reduced for
large $M$ but can be recovered at higher values of illumination
$\mathcal{I}$. A discussion about that point can be found in the
Appendix after Eq. (\ref{Ncrosscorr}). This means that  ThGI
performs best in the single-mode regime, while  TwGI,  even at low
brightness, can be easily brought to a high-illumination level by
accumulating a large number of temporal modes in a single frame,
without any decrease in the SNR. However, provided that the source
has the same parameters, i.e. the same brightness $\mu$, the same
$M$ and therefore the same $\mathcal{I}$, the advantages of
quantum light can be of several orders of magnitude, even for
large illumination $\mathcal{I}$.

In conclusion, our discussion demonstrates that single-mode
($M=1$)  ThGI  is the best solution, considering the less demanding
experimental resources, when one is not limited in the brightness
of light. On the other hand, in situations where a low light
level, $\mathcal{\mu}<1$, is needed,  for instance because of the
photosensitivity of the object, quantum light provides a larger
SNR (even for a  relatively large illumination level).

\section{The experiment}\label{experiment}

For the experimental comparison of GI obtained with thermal light and twin beams, we
used two setups.  In the first case, GI was performed with bright
pseudo-thermal radiation (more than  $10^4$  photons per radiation
mode). In the second case, nonclassical radiation was used,
namely, squeezed vacuum obtained via type-II parametric
down-conversion.

In both experiments, we used the `numerical mask' method suggested
in Ref.~\cite{Basano}. We  selected two regions, one in the object
beam, where we virtually inserted the mask, and the other in the
reference beam. In the bucket channel, we simulated the presence of
a completely opaque mask, which fully transmitted light within a
certain area $\mathcal{A}_{in}$ and fully absorbed light outside
it. This was done by simply setting to zero the signals of all
pixels outside of this area. This method, which is only possible
with an array photodetector in the bucket channel, presents the
advantage of arbitrarily varying the shape and size of the mask
without any technical difficulties and provides an ideal limit for
this part of the set-up in a realistic situation.

\subsection{Experimental  setup with pseudo-thermal light}\label{thermal_exp}

\begin{figure}[tbp]
\center
\includegraphics[width=10cm, height=7cm ]{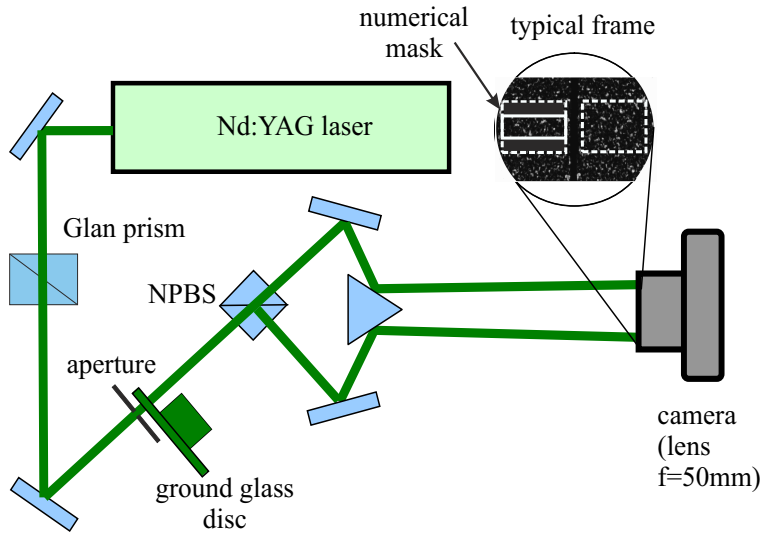}
\caption{Experimental setup for realizing ghost imaging with pseudo-thermal light.
}\label{setup_th}
\end{figure}

The experiment with pseudo-thermal light ghost imaging was performed
with the setup shown in Fig.~\ref{setup_th}
 \cite{Agafonov}. Second-harmonic
radiation of a Nd:YAG laser with the wavelength $532$ nm, pulse
duration $10$ ns, and repetition rate $47$ Hz was incident on an
Arecchi's rotating ground-glass disc generating pseudo-thermal light
\cite{Ar}. The radiation scattered by the disc was split by a
non-polarizing beamsplitter and then both output beams were
registered by two different parts of the matrix of a commercial
digital photographic camera Sigma SD14. The far-field speckle
patterns, symmetrical in the two parts, were focused on the matrix
by a lens with the focal length $50$ mm.  As a result, the typical
speckle size on the matrix was about $0.230$ mm. With the pixel size
of the matrix being $7.8\,\,\mu$m, a single speckle occupied, on the
average, an area of $30\times30$ pixels. This was confirmed by the
measurement of the spatial intensity correlation function.

The rotation rate of the disc was chosen in such a way that the
speckle pattern did not change noticeably during a single pulse but
changed completely from pulse to pulse. The camera was taking frames
with exposure times $1/50$ s, which was much larger than the pulse
duration but still less than the distance between the pulses. As a
result, most of the frames contained the speckle pattern from a
single pulse~(\cite{Agafonov}, Fig.~\ref{setup_th}) and few were
`empty'. The total number of captured frames was $5000$. In each
frame, we selected square regions in the two symmetrical parts
~(Fig. 6 of Ref.~\cite{Agafonov}), further referred to as the object
and reference channels. As before, in the object channel, the
'masks' were introduced numerically, by simply ignoring the readings
of the pixels outside them.

The intensity of the laser light was reduced by using a Glan prism;
by measuring the signals from separate pixels of the camera versus
the input intensity we verified that the camera was operating in a
linear regime and was not saturated. It is important that the output
data of the camera was written in the RAW format. The mean values of
signals from separate pixels were on the order of $300$, which
indicates that the number of incident photons was at least as much
as that. Hence,  per single speckle, i.e. for a single radiation
mode, we had photon numbers as high as $3\cdot10^4$. As the pulse
duration was much smaller than the typical time of the speckle
pattern fluctuation, and the pixel size was much smaller than the
speckle size, the number of modes was $M=1$.  The
transmission/quantum efficiency was quite low (less than 0.01), due
to the beam attenuation, but, as discussed in section II, this is
not important for ThGI.

\subsection{\ Experimental setup with type-II PDC}\label{PDC_exp}

\begin{figure}[tbp]
\center
\includegraphics[width=10cm, height=7cm ]{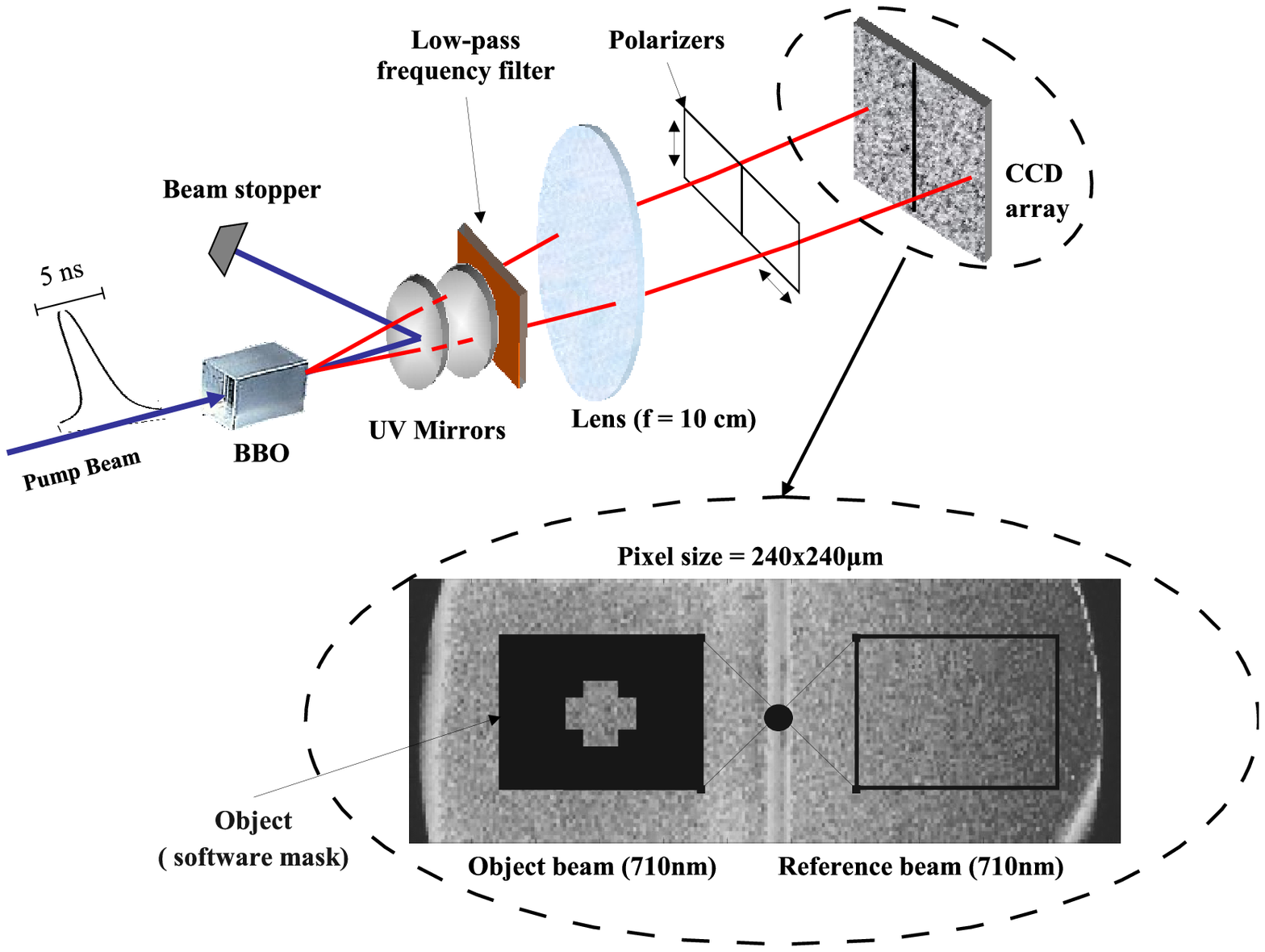}
\caption{Scheme of the  setup for realizing ghost imaging with twin
beams produced by Type II parametric down conversion.
}\label{Set_up-PDC}
\end{figure}

In contrast to most ghost imaging experiments reported in the
literature that are realized with faint PDC, here we set an
experiment in the mesoscopic regime using a mesoscopic PDC source
and  a CCD  camera as a photon number resolving detector. This
represents an important step in view of practical applications of
this method. The setup is reported in Fig.\ref{Set_up-PDC}.

 For generating the twin beams we used a type II
BBO non-linear crystal ($l=7$ mm) pumped by the third harmonic (355
nm) of a Q-switched Nd:YAG  laser. The pulses had a duration of
$\mathcal{T}_p=5$ ns with a repetition rate of 10 Hz and a maximum
energy, at the selected wavelength, of about 200 mJ. The spurious
non-Gaussian  components were eliminated by spatial filtering (a
lens with a focal length of 50 cm and a diamond pin-hole, 250 $\mu$m
of diameter) of the beam that was re-collimated with a diameter of
$w_{p}=1.25$ mm at the crystal. After the BBO, the pump was stopped
by two UV mirrors (with $\simeq 98\%$ declared transmission at 710
nm), and by a low-pass frequency filter ($\simeq 95\%$ transmission
at 710 nm). The orthogonally polarized down-converted signal and
idler beams were separated by two polarizers ($\simeq 97\%$
transmission) and finally the far field was registered by a 1340X400
CCD (Princeton Pixis:400BR  with pixel size of 20 $\mu$m, quantum
efficiency of 80\% and  readout noise  $4$ electrons/pixel). A
mechanical shutter provided the synchronization with the laser so
that each acquired frame corresponded to the PDC emission generated
by a single shot of the laser. The far field was observed at the
focal plane of the lens with 10 cm focus in  an '$f-f$' optical
configuration. Our setup was developed exploiting the potentiality
of the quantum scheme \cite{BridaNPHOT2010},  which, according to
the theory, should outperform the classical light scheme for the
same brightness (number of photon per spatio--temporal mode). For
optimizing the detection of quantum correlations, basically two
requirements should be fulfilled. First, transmission/detection and
collection efficiency must be large. We realized it by reducing the
optical losses after the crystal and using a CCD camera with high
quantum efficiency, while the collection efficiency can be kept high
if the pixel area $\mathcal{A}_{pix}$ is larger than the coherence
area $\mathcal{A}_{coh}$ (approximatively a speckle), i.e. a pixel
should collect more than one spatial mode \cite{sp}. The second
point is that the detection noise must be low compared with the shot
noise. For both these reasons, it is very  efficient to perform
hardware binning of the physical pixels. It consists  of grouping
the physical pixels in squared blocks, each of them being processed
by the CCD electronics as single "superpixel". The number of photons
collected by a superpixel is the sum of photon numbers of each
pixel, whereas the  readout noise is just slightly increased with
respect to the one of a  single pixel. We chose its dimension
comparable to $\mathcal{A}_{coh}$, i.e. the pixel (hereinafter we
discard the prefix "super") size was set to $240\times240(\mu
m)^{2}$. The "illumination" $\mathcal{I}$, defined as  the number of
photons detected by  a pixel  in a single shot image (frame), was
about $\mathcal{I}=1900$. The expected number of temporal modes
$\mathcal{M}_{t}=\mathcal{T}_{p} / \mathcal{T}_{coh}$ detected in
one frame was $5\cdot 10^{3}$, considering the coherence time
$\mathcal{T}_{coh}$ of PDC around one picosecond. The number of
spatial modes detected by the pixel was
$\mathcal{M}_{sp}=\mathcal{A}_{pix}/\mathcal{A}_{coh}\simeq4$. A
measure of the coherence area, basically determined by the pump
transverse diameter, is the size of the spatial cross-correlation of
the signal and idler intensity patterns that has been evaluated to
$\mathcal{A}_{coh}\sim 120\times120(\mu m)^{2}$. The total number of
modes $M=\mathcal{M}_{t}\times\mathcal{M}_{sp}$ turned out to be
compatible with the level of excess noise $\mathcal{E} \equiv
\mathcal{I}/M\sim 0.12$ due to the thermal statistics in the single
beam. Thus, the number of photons per single space-temporal modes
was $\mu=\mathcal{I}/(\eta M)\simeq0.20$. Under these conditions, we
recorded $4000$ frames.

\subsection{Experimental Results}\label{Results}
After having discarded all the frames corrupted by incoming cosmic
rays,  as well as `empty' frames in the  ThGI  setup,  we started
the analysis  by defining two correlated regions in the object and
in the reference arm.  In the  TwGI  setup, the regions included
$13\times15=195$ pixels, while in the  ThGI  setup, they
contained $750\times750$ pixels. These regions were the same for all the
frames.

In the twin beams set-up, one of the main problems  we had to cope
with was a strong instability of the Q-switched laser power from
pulse to pulse (about 14\%). Since the expectation value of the
number of photons per mode $\mu$ is proportional to the square of
the power $P$ of the pump
($\mu\propto\sinh^2(\mbox{const}\sqrt{P})$), fluctuations of $P$
lead to the fluctuations of $\mu$. As a consequence, the temporal
statistics on many pulses is characterized by  the mean $\bar{\mu}$
and the variance $V(\mu$).  Indeed, using the equations presented in
the Appendix, we have obtained the expectation value of $Cov$ inside
and outside of the mask taking into account the pump instability:
\begin{equation}\label{}
Cov'_{in}=\eta^2M[\bar{\mu}(1+\bar{\mu})+V(\mu)(1+RM)],
\end{equation}

\begin{equation}\label{}
Cov'_{out}=\eta^2M^2V(\mu)R.
\end{equation}

When $M$ and $R$ are large, as in our experiment, the fluctuations
of the pump dominate and this introduces a  nonzero covariance for
pixels in the reference arm that are, in principle, uncorrelated
with the mask. The overall result is a strong noise of "background
correlation" that hides the PDC spatial correlation, essential in
the ghost image reconstruction. As a matter of fact, this effect
would lead to an increased noise on the ghost image and a consequent
decreasing of SNR.

In order to compensate for the instability of the pump, we
normalized each frame by its average value taken over  a certain CCD
spatial region.
This procedure completely removed the
problem.

\begin{figure}[h]
\center
\includegraphics[width=10cm, height=7cm ]{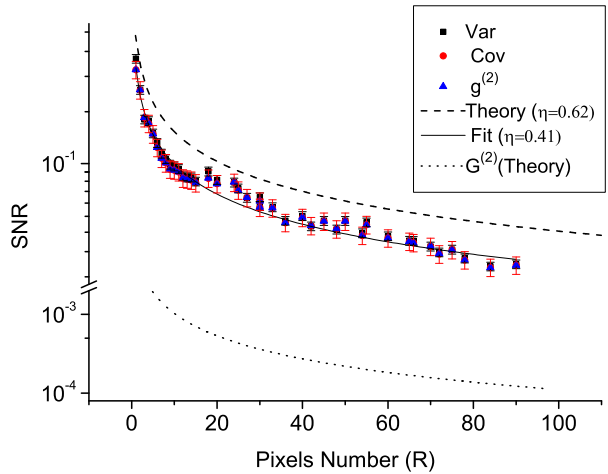}
\caption{ Signal-to-noise ratio for TwGI as a  function of the
resolution parameter $R$. The three series of experimental data
refer to the different correlation parameters considered in our
analysis $g^{2}$, $Cov$, $Var$. They completely overlap in this
regime, as well as the theoretical curves (dashed line). The
theoretical curve for $G^{2}(R)$ is also reported (dotted line).
The SNR is normalized by the square root of the number of frames
of the sample $\mathcal{K}=4000$. }\label{SNRvsRpdc}
\end{figure}

In Fig.\ref{SNRvsRpdc} we present the  obtained experimental values
of the SNR versus the resolution parameter $R$  of the mask for the
case of TwGI. The data set is in agreement with the theory where the
experimental  parameters $M$,  $\mu$ and $\eta$ have been estimated
independently. However, the best fit is obtained when $\eta=0.42$
that is smaller than the  value 0.62 that was measured very
accurately in \cite{BDGRR-OE-2010}. This discrepancy can be
explained by the relatively small pixel size (reduced collection
efficiency of correlated photons) and the presence of background
noise generated by detection and straylight. A detailed analysis of
such problems can be found in \cite{BDGRR-OE-2010}. We stress that
in our experiment we exploited successfully the theoretical results
pointing out that the multimode regime is the most appropriate for
TwGI.  In fact, by collecting
$M\sim\mathcal{M}_{sp}\cdot\mathcal{M}_{t}=4\cdot5000$ we reached a
level of illumination $\mathcal{I}=1900$ such that the experimental
sources of noise (like electronic read-noise of the CCD and
straylight ecc.) were almost negligible. For $G^{2}$ we only report
the theoretical prediction, which is close to zero. In accordance to
this prediction, our experimental data involving 4000 frames was
insufficient to produce any ghost image with $G^{2}$ protocol.

 In summary, we have demonstrated ghost imaging with twin beams in a
mesoscopic regime ($\mu=0.2$,
$M\equiv\mathcal{M}_{sp}\cdot\mathcal{M}_{t}=4\cdot5000$,
$\mathcal{I}=1900$) by a photon-number resolving detector for such
parameters of the source (number of photons per mode and the
illumination level) for which thermal ghost imaging provides much
worse SNR, according to the values reported in Fig. \ref{SNRvsI-M}.

\begin{figure}[tbp]
\center
\includegraphics[width=10cm, height=7cm ]{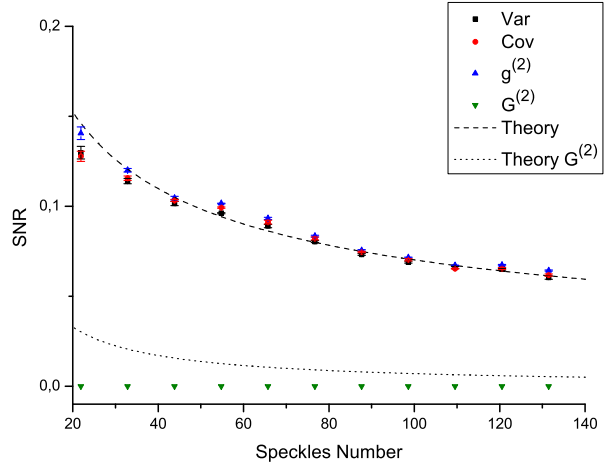}
\caption{ Signal-to-noise ratio for ThGI as a  function of the
resolution parameter $R$. The three series of experimental data
refer to the different correlation parameters considered in our
analysis $g^{2}$, $Cov$, $Var$. They completely overlap in this
regime, as well as the theoretical curves (dashed line). The
theoretical curve for $G^{2}(R)$ is also reported (dotted line)  but
in experiment, no readable image could be obtained from our dataset.
 The SNR is normalized by the square root of the number of
frames of the sample. }\label{AllTogether}
\end{figure}

On the other hand, in figure~\ref{AllTogether} shows the results
of ThGI experiment. The illumination level was about
$\mathcal{I}\approx 300$, photon number per mode $\mu\approx 10^4$,
the number of modes $M=1$, and the resolution parameter $R$ was
varied from $R\approx4$ to $R\approx200$ by changing the size of
rectangular 'numerical masks'. One can see that the experimental
points agree well with the theoretical dependence for SNR, which,
according to Table~\ref{table}, is in this case similar for all
three methods (based on variance, covariance, and $g^{(2)}$) and
given by $SNR=\sqrt{2R}$.

\section{Conclusions}\label{Conclusion}

For providing the reader, in view of practical  applications, with a
general exhaustive description of ghost imaging, we have discussed
both thermal and  twin-beam  GI for four different
 protocols, based on the measurement of the following values:
the Glauber correlation function $G^{(2)}=\left\langle
\mathbb{N}_{1}N_{2}\right\rangle$, the normalized Glauber's
correlation function $g^{(2)}=G^{(2)}/
\langle\mathbb{N}_{1}\rangle\langle N_{2}\rangle$,  the covariance
$Cov=\left\langle \delta\mathbb{N}_{1}\delta N_{2}\right\rangle$,
and the variance of the difference signal  $Var=\langle\delta^{2}
(\mathbb{N}_{1}-N_{2})\rangle$.

A first significant result, in accordance with the analysis in
\cite{Chan2010} for thermal light, is that the $G^{(2)}$ method
performs much worse than the others (the SNR falling rapidly with
the resolution parameter, i.e. the number of speckles transmitted
through the mask in the object channel, and with the number of
spatio--temporal modes detected by the reference detector): thus,
albeit often considered in the literature for the sake of
simplicity,  the $G^{(2)}$ method is not worth using in real
applications.

On the other hand, the other methods have similar performances. For
all of them, twin-beam GI performs largely better than thermal-light
GI at low illumination levels (few photons per pixel in a single
shot), while they become equivalent at high illumination levels
(many photons per pixel in a single shot). Furthermore, TwGI is
insensitive to the number of modes collected by a pixel and can be
brought to a large illumination level by accumulating a large number
of temporal modes in a single shot, even maintaining low brightness,
 which  means low photon number per spatio--temporal
mode. This is what is realized in our experimental demonstration, in
which about 20000 modes are accumulated in the 5 ns duration of the
pump pulse. ThGI performs best in the single-mode regime. In
summary, ThGI is preferable whenever one does not have limits on the
brightness (e.g. photo--sensible samples), due to its simpler
realization.

Finally, we have demonstrated the perfect agreement of the
developed general theory of GI with experiment in some
particularly interesting (and sometimes unexplored) regimes for
both TwGI and ThGI.

\section{Appendix}\label{App}

In this Appendix we calculate the correlation functions needed for
evaluating the SNR of different GI schemes.

 Both for thermal and PDC
light we first consider the single-mode statistics and then we
extend it to a more general multi-mode case. Only later we
introduce the bucket detector that integrates the light over all
the spatial modes in the `mask' (`object') beam.

PDC \cite{Ben}, in the approximation of a plane and monochromatic
pump beam of frequency $\omega_{p}$, can be described by an
evolution operator $\mathbb{U}$ that is the product of
independent operators, each acting on a couple of symmetric modes
$(\mathbf{q},\Omega)$ and $(-\mathbf{q},-\Omega)$
\cite{RuoASL2009}:

\begin{equation}\label{U}
\mathcal{\mathbb{U}}=\exp\left(-\frac{i}{\hbar}\int_{0}^{\infty}H(t')
dt'\right) =\bigotimes_{\mathbf{q},\Omega}\exp\left[ \xi
a_{1}^{\dagger}(\mathbf{q},\Omega)a_{2}^{\dagger}(-\mathbf{q},-\Omega)-h.c.\right],
\end{equation}
where $\pm\Omega$ is the detuning of  the signal(1) and idler(2)
photons with respect to the degeneracy frequency $\omega_{p}/2$,
$\pm\mathbf{q}$ are the symmetric and correlated transverse
wavevectors, $a_{1}^{\dagger}$ and $a_{2}^{\dagger}$ are the creation operators of
the signal and idler photons, respectively, and
$\xi(\mathbf{q},\Omega)$ is the coupling coefficient. Therefore,
it can be seen as a collection of independent processes where
couples of symmetric modes are correlated and the output field
operators for signal and idler  are
\begin{eqnarray}\label{input-output}
\nonumber
b_{1}(\mathbf{q},\Omega)=U(\xi)a_{1}(\mathbf{q},\Omega)+V(\xi)a^{\dagger}_{2}(-\mathbf{q},-\Omega),\\
b_{2}(\mathbf{q},\Omega)=U(\xi)a_{2}(\mathbf{q},\Omega)+V(\xi)a^{\dagger}_{1}(-\mathbf{q},-\Omega),
\end{eqnarray}
where $U(\xi)=\cosh(\xi)$ and $V(\xi)=\sinh(\xi)$ determine
the gain of the process by their maximum value and the coherence
time by their inverse width in $\Omega$. Here the phases are not
considered because they do not play any role in the photon-number
statistics.

In our model we assume $\eta_{j}$ ($j=1,2$) to be the detection
probabilities for the photons in the mode $j$. Here, $\eta_{j}$ is a
coefficient that includes transmission, detection (quantum
efficiency, QE, of the detector) and collection efficiency $\eta_{j,
coll}$ of the photon emitted in a certain spatial modes. In
particular, in our model the collection efficiency can be less than
one when the pixel of the spatially resolving detector is smaller
than the spatial mode at the detection plane. In that case, the
probability to collect a photon emitted in a certain mode is
$\eta_{j, coll}= \mathcal{A}_{pix}/\mathcal{A}_{coh}$. All the
losses can be taken into account by a model in which a beam splitter
with transmission $\eta_{j}$ is preceding an ideal detector (QE$=1$)
\cite{mat}. For the calculation of the photon-number correlation
function $n_{j}=b^{\dag}_{j}b_{j}$, this leads to a simple
substitution $b_{j}\rightarrow\sqrt{\eta_{j}} b_{j}$ in the
normally ordered expressions for the operators. The single-mode photon-number
 statistics as well as the two-mode correlation can be
obtained through a straightforward calculation by the relations
(\ref{input-output}), with the input state being the vacuum one:

\begin{eqnarray}\label{n-corr}
\langle n_{j}\rangle     &=& \eta_{j}\mu ,
\\\nonumber \langle n_{j}^{2}\rangle &=& \eta_{j}\mu
+2\eta_{j}^{2}\mu^{2},\\\nonumber \langle n_{j}^{3}\rangle &=& \mu  \eta_{j} \left(1+6 \mu  \eta_{j}+6 \mu ^2
\eta_{j}^2\right),\\\nonumber \langle n_{j}^{4}\rangle &=& \mu \eta_{j} \left(1+14 \mu \eta_{j}+36 \mu ^2
\eta_{j}^2+24 \mu ^3 \eta_{j}^3\right),\\\nonumber
\end{eqnarray}
\begin{eqnarray}
\label{n-crosscorr}
\langle n_{1}n_{2}\rangle   &=& \mu  (1+2 \mu ) \eta_{1}
\eta_{2},\\\nonumber \langle n_{1}^{2}n_{2}\rangle &=& \mu
\eta_{1} \eta_{2} \left(1+6 \mu ^2 \eta_{1}+\mu  (2+4
\eta_{1})\right),\\\nonumber \langle n_{1}^{2}n_{2}^{2}\rangle &=&
\mu  \eta_{1} \eta_{2} \left[1+24 \mu ^3 \eta_{1} \eta_{2}+6 \mu
^2 (\eta_{1}+\eta_{2}+4 \eta_{1} \eta_{2})+2\mu  (1+2 \eta_{1}+2
\eta_{2} +2\eta_{1}\eta_{2})\right],\\\nonumber \langle
n_{1}^{3}n_{2}\rangle    &=& \mu  \eta_{1} \eta_{2}\left[1+24 \mu
^3 \eta_{1}^2+18 \mu ^2 \eta_{1} (1+\eta_{1})+2 \mu  (1+6
\eta_{1})\right]. \\\nonumber
\end{eqnarray}
For our purposes we focus on the correlation in the far field
obtained as the focal plane of a thin lens of focal length $f$ in an
$f-f$ configuration and around degeneracy ($\Omega=0$). Here, any
transverse mode $\mathbf{q}$ is associated with a single position
$\mathbf{x}$ in the detection (focal) plane according to the
geometric transformation $(2c f/\omega_{p})\mathbf{q}\rightarrow
\mathbf{x}$ \cite{th2}, with $c$ being the speed of light. We
observe that the diffraction effects arising from the transverse
finite dimension of the optical system
 (typically the pump profile in the twin beam generation process) lead to a nonzero transverse coherence area, $\mathcal{A}_{coh}$,
 in the far field, representing the uncertainty in the arrival point of the correlated photons.
The fluctuation of the difference photon number is
$\langle\delta^{2}(n_{1}-n_{2})\rangle_{TW}\equiv\langle\delta^{2}n_{1}\rangle
+\langle\delta^{2}n_{2}\rangle-2\langle\delta n_{1}\delta
n_{2}\rangle=2\eta \mu (1-\eta)$ for $\eta=\eta_{1}=\eta_{2}$. Thus,
in the ideal lossless case ($\eta=1$), a perfect correlation appears
in the photon number since
$\langle\delta^{2}(n_{1}-n_{2})\rangle_{TW}=0$ (sub-shot-noise
regime \cite{BridaPRL2009,BridaNPHOT2010}).

In the case of thermal radiation we assume a set of independent
modes $a_{i}$ (i=1,..) addressed to the input of a 50\% beam
splitter performing the transformations
$b_{1,i}=(a_{i}+a_{vac,i})/\sqrt{2}$ and
$b_{2,i}=(a_{i}-a_{vac,i})/\sqrt{2}$, where $a_{vac}$ is the mode
at the vacuum input port of the beam splitter.  The
photon-number correlation of the beams at the two output ports are
obtained by exploiting the rule $\langle
(a_{i}^{\dag})^{p}a_{i}^{p}\rangle=p!\langle
a_{i}^{\dag}a_{i}\rangle^{p}$, for normally ordered moments
 of thermal light. For the single-beam moments, taking
into account the losses, we found again the expressions in
Eqs.~(\ref{n-corr}). This is in agreement with the fact that each
beam from PDC, considered separately, manifests thermal
statistics. On the other hand, the correlations of photon numbers
between thermal beams 1 and 2 are

\begin{eqnarray}\label{n-crosscorrth}
\langle n_{1}n_{2}\rangle   &=& 2 \mu^{2} \eta_{1}\eta_{2},\\
\nonumber
\langle n_{1}^{2}n_{2}\rangle    &=& 2\mu^{2}\eta_{1} \eta_{2} \left(1+3 \mu \eta_{1}\right),\\
\nonumber
\langle n_{1}^{2}n_{2}^{2}\rangle    &=& 2\mu^{2}  \eta_{1}\eta_{2} \left[1+12 \mu^{2} \eta_{1} \eta_{2}+3 \mu
(\eta_{1}+\eta_{2})\right],\\
\nonumber \langle n_{1}^{3}n_{2}\rangle &=& 2\mu  \eta_{1} \eta_{2}\left[1+9
\eta_{1}\mu +12\mu^{2}\eta_{1}^{2}\right]. \\
\nonumber
\end{eqnarray}
In contrast to the quantum case, the correlation between
the photon numbers in the coupled modes is not perfect; hence, for
thermal light we have
$\langle\delta^{2}(n_{1}-n_{2})\rangle_{Th}=2\eta \mu$, which
represents the shot noise limit. At the same time, the excess
noise $\sim\mu^{2}$ in the second equation in (\ref{n-corr})
disappears regardless of the losses.

According to our experimental realization (but without loss of
generality), we start by considering a situation in which both
reference and object beams are detected by a spatially resolving
array of detectors (the pixels of a CCD camera in our case). Each pixel
collects a number $M\geq1$ of independent spatio-temporal modes in the
far field and the total number of photons detected by the pixel is
$N_{j}=\Sigma_{i=1}^{M}n_{j,i}$, with $j=1,2$. It is easy to show
that the moments of the $N$ distribution can be expressed in terms
of single-mode moments in Eqs.~(\ref{n-corr}) as

\begin{eqnarray}\label{Ncorr}
 \small
  \left\langle N_{j}\right\rangle &=& M\left\langle n_{j}\right\rangle,\\\nonumber
  \left\langle N_{j}^{2}\right\rangle &=& M \left\langle n_{j}^{2}\right\rangle+M (M-1)\left\langle n_{j}\right\rangle^{2},\\\nonumber \left\langle N_{j}^{3}\right\rangle &=& M \left\langle n_{j}^{3}\right\rangle+3M (M-1)\left\langle n_{j}^{2}\right\rangle\left\langle n_{j}\right\rangle+M(M-1)(M-2)\left\langle n_{j}\right\rangle^{3},\\\nonumber
  \left\langle N_{j}^{4}\right\rangle &=& M \left\langle n_{j}^{4}\right\rangle+M (M-1)\left(3\left\langle n_{j}^{2}\right\rangle^{2}+4\left\langle n_{j}^{3}\right\rangle\left\langle n_{j}\right\rangle\right)+6M(M-1)(M-2)\left\langle n_{j}^{2}\right\rangle\left\langle n_{j}\right\rangle^{2}+\\\nonumber &+&M(M-1)(M-2)(M-3)\left\langle n_{j}\right\rangle^{4}.\nonumber
\end{eqnarray}
Since the modes of signal and idler fields are pairwise
correlated, the correlators of the count numbers in two
symmetrical pixels (collecting $M$ correlated modes) are

\begin{eqnarray}\label{Ncrosscorr}
  \left\langle N_{1}N_{2}\right\rangle &=& M\left\langle n_{1}n_{2}\right\rangle+M(M-1)\left\langle n_{1}\right\rangle \left\langle n_{2}\right\rangle,\\\nonumber
  \left\langle N_{1}^{2}N_{2}\right\rangle &=& M\left\langle n_{1}^{2}n_{2}\right\rangle+M (M-1)\left(\left\langle n_{1}^{2}\right\rangle \left\langle n_{2}\right\rangle+2\left\langle n_{1}n_{2}\right\rangle\left\langle n_{1}\right\rangle\right)\\\nonumber
  &+&M(M-1)(M-2)\left\langle n_{1}\right\rangle^{2}\left\langle n_{2}\right\rangle,\\\nonumber
  \left\langle N_{1}^{2}N_{2}^{2}\right\rangle &=&M\left\langle n_{1}^{2}n_{2}^{2}\right\rangle\\\nonumber
  &+& M(M-1)\left(\left\langle n_{1}^{2}\right\rangle\left\langle n_{2}^{2}\right\rangle+2\left\langle n_{1}^{2}n_{2}\right\rangle \left\langle n_{2}\right\rangle+2\left\langle n_{1}n_{2}\right\rangle^{2}+2\left\langle n_{1}n_{2}^{2}\right\rangle\left\langle n_{1}\right\rangle\right)\\\nonumber
  &+&M(M-1)(M-2)\left(\left\langle n_{1}^{2}\right\rangle\left\langle n_{2}\right\rangle^{2}+\left\langle n_{1}\right\rangle^{2} \left\langle n_{2}^{2}\right\rangle+4\left\langle n_{1}n_{2}\right\rangle\left\langle n_{1}\right\rangle\left\langle n_{2}\right\rangle\right)\\\nonumber
  &+&M(M-1)(M-2)(M-3)\left\langle n_{1}\right\rangle^{2}\left\langle n_{2}\right\rangle^{2},\\\nonumber
  \left\langle N_{1}^{3}N_{2}\right\rangle &=&M\left\langle n_{1}^{3}n_{2}\right\rangle\\\nonumber
  &+& M(M-1)\left(\left\langle n_{1}^{3}\right\rangle\left\langle n_{2}\right\rangle+3\left\langle n_{1}^{2}\right\rangle \left\langle n_{1}n_{2}\right\rangle+3\left\langle n_{1}^{2}n_{2}\right\rangle\left\langle n_{1}\right\rangle\right)\\\nonumber
  &+&M(M-1)(M-2)\left(3\left\langle n_{1}n_{2}\right\rangle\left\langle n_{1}\right\rangle^{2}+3\left\langle n_{1}^{2}\right\rangle\left\langle n_{1}\right\rangle\left\langle n_{2}\right\rangle\right)\\\nonumber
  &+&M(M-1)(M-2)(M-3)\left\langle n_{1}\right\rangle^{3}\left\langle
  n_{2}\right\rangle\nonumber.
\end{eqnarray}

The correlators $\left\langle N_{1}N_{2}^{2}\right\rangle$ and
$\left\langle N_{1}N_{2}^{3}\right\rangle$ can be obtained by simply
exchanging the indexes 1 and 2 in the second and fourth equations of
(\ref{Ncrosscorr}).  From the first lines of Eqs.~(\ref{Ncorr}) and
(\ref{Ncrosscorr}) we see that $\langle\delta N_{1}\delta
N_{2}\rangle=M \langle\delta n_{1}\delta n_{2}\rangle$
($\eta_{1}=\eta_{2}=1$). By substituting the two-mode correlators
following from Eqs.~(\ref{n-corr}) and (\ref{n-crosscorrth}) we have
$\langle\delta N_{1}\delta N_{2}\rangle_{TW}=\langle
N_{1}\rangle+\langle N^{2}_{1}\rangle/M$ and $\langle\delta
N_{1}\delta N_{2}\rangle_{TH}=\langle N^{2}_{1}\rangle/M$ for
quantum and  thermal light,  respectively. These expressions show
that  thermal  correlations are sensitive to the  numbers  of modes
in both channels, while the shot noise term is not reduced by the
increase in $M$. This provides an intuitive explanation of the
different behavior between  TwGI and ThGI  with respect to the
number of modes $M$.

Now we introduce the bucket detector whose output is represented by
the operator $\mathbb{N}_{1}=\Sigma_{k=1}^{R}N_{k,1}$, i.e. a sum of
the photon numbers over $R$ independent sets of spatial cells, each
one collecting $M\geq1$ modes. $R$ coincides with the number of
pixels of the ghost image if the pixel is sufficiently large to
collect more than one spatial mode, otherwise it coincides with the
number of spatial modes (speckles) itself.
\begin{eqnarray}\label{N|corr}
 \small
  \left\langle \mathbb{N}_{1}\right\rangle &=& R\left\langle N_{1}\right\rangle,\\\nonumber
  \left\langle \mathbb{N}_{1}^{2}\right\rangle &=& R \left\langle N_{1}^{2}\right\rangle+R (R-1)\left\langle N_{1}\right\rangle^{2},\\\nonumber \left\langle \mathbb{N}_{1}^{3}\right\rangle &=& R \left\langle N_{1}^{3}\right\rangle+3R (R-1)\left\langle N_{1}^{2}\right\rangle\left\langle N_{1}\right\rangle+R(R-1)(R-2)\left\langle N_{1}\right\rangle^{3},\\\nonumber
  \left\langle \mathbb{N}_{1}^{4}\right\rangle &=& R \left\langle N_{1}^{4}\right\rangle+R (R-1)\left(3\left\langle N_{1}^{2}\right\rangle^{2}+4\left\langle N_{1}^{3}\right\rangle\left\langle N_{1}\right\rangle\right)+6R(R-1)(R-2)\left\langle N_{1}^{2}\right\rangle\left\langle N_{1}\right\rangle^{2}+\\\nonumber &+&R(R-1)(R-2)(R-3)\left\langle N_{1}\right\rangle^{4}.\nonumber
\end{eqnarray}
The correlators between the readings of the bucket detector and
the ones of an arbitrary pixel in the reference channel
registering light correlated with the one passing through the mask
(dubbed with the subscript "in") are

\begin{eqnarray}\label{N|crosscorr}
  \left\langle \mathbb{N}_{1}N_{2,in}\right\rangle &=& \left\langle N_{1}N_{2}\right\rangle+(R-1)\left\langle N_{1}\right\rangle \left\langle N_{2}\right\rangle,\\\nonumber
  \left\langle \mathbb{N}_{1}^{2}N_{2,in}\right\rangle &=& \left\langle N_{1}^{2}N_{2}\right\rangle+ (R-1)\left(\left\langle N_{1}^{2}\right\rangle \left\langle N_{2}\right\rangle+2\left\langle N_{1}N_{2}\right\rangle\left\langle N_{1}\right\rangle\right)\\\nonumber
  &+&(R-1)(R-2)\left\langle N_{1}\right\rangle^{2}\left\langle N_{2}\right\rangle,\\\nonumber
  \left\langle \mathbb{N}_{1}N_{2,in}^{2}\right\rangle &=& \left\langle N_{1}N_{2}^{2}\right\rangle+ (R-1)\left\langle N_{1}\right\rangle \left\langle N_{2}^{2}\right\rangle,\\\nonumber
  \left\langle \mathbb{N}_{1}^{2}N_{2,in}^{2}\right\rangle &=&\left\langle N_{1}^{2}N_{2}^{2}\right\rangle\\\nonumber
  &+& (R-1)\left(\left\langle N_{1}^{2}\right\rangle\left\langle N_{2}^{2}\right\rangle+2\left\langle N_{1}N_{2}^{2}\right\rangle \left\langle N_{1}\right\rangle\right)\\\nonumber
  &+&(R-1)(R-2)\left\langle N_{1}\right\rangle^{2}\left\langle N_{2}^{2}\right\rangle,\\\nonumber
  \left\langle \mathbb{N}_{1}^{3}N_{2,in}\right\rangle &=&\left\langle N_{1}^{3}N_{2}\right\rangle\\\nonumber
  &+& (R-1)\left(\left\langle N_{1}^{3}\right\rangle\left\langle N_{2}\right\rangle+3\left\langle N_{1}^{2}\right\rangle \left\langle N_{1}N_{2}\right\rangle+3\left\langle N_{1}^{2}N_{2}\right\rangle\left\langle N_{1}\right\rangle\right)\\\nonumber
  &+&(R-1)(R-2)\left(3\left\langle N_{1}N_{2}\right\rangle\left\langle N_{1}\right\rangle^{2}+3\left\langle N_{1}^{2}\right\rangle\left\langle N_{1}\right\rangle\left\langle N_{2}\right\rangle\right)\\\nonumber
  &+&(R-1)(R-2)(R-3)\left\langle N_{1}\right\rangle^{3}\left\langle N_{2}\right\rangle,\\\nonumber
  \left\langle \mathbb{N}_{1}N_{2,in}^{3}\right\rangle&=&\left\langle N_{1}N_{2}^{3}\right\rangle+(R-1)\left\langle N_{1}\right\rangle\left\langle
  N_{2}\right\rangle^{3}\nonumber.
\end{eqnarray}
The analogous equation for the "out" case is obtained by  using  the statistical independence of $\mathbb{N}_{1}$ and $N_{2,out}$, i.e. by  substituting  $\langle\mathbb{N}_{1}^{p}N_{2,out}^{q}\rangle=\langle \mathbb{N}_{1}^{p}\rangle\langle N_{2,out}^{q}\rangle$, and then applying  Eqs.~(\ref{N|corr})  and (\ref{Ncorr}).

By using the definition (\ref{SNRdef}) and Eqs.~(\ref{N|corr}),
(\ref{N|crosscorr}) we arrive to our full analytical expressions
for the SNR for the protocols considered in the paper.

\section{Acknowledgements} This work has been supported by Regione Piemonte.
We thank Ivan Agafonov for providing the set of frames with
thermal-light speckle patterns.

\end{document}